\documentclass[preprint,showpacs,aps,prb,showpacs]{revtex4}

\usepackage{graphicx}
\usepackage{dcolumn}

\begin{document}

\bibliographystyle{apsrev}


\title{Energetics of intrinsic point defects in ZrSiO$_4$}

\author{J.M. Pruneda}
\author{Emilio Artacho}
\affiliation{Department of Earth Sciences, University of Cambridge,
Downing Street, Cambridge, CB2 3EQ, United Kingdom}

\date{\today}

\begin{abstract}
Using first principles calculations we have studied the formation 
energies, electron and hole affinities, and electronic levels of 
intrinsic point defects in zircon.  The atomic structures of charged 
interstitials, vacancies, Frenkel pairs and anti-site defects are 
obtained.  The limit of high concentration 
of point defects, relevant for the use of this material in nuclear 
waste immobilization, was studied with a variable lattice relaxation 
that can simulate the swelling induced by radiation damage.  
The limit of low concentration of defects is simulated
with larger cells and fixed lattice parameters.   Using known band
offset values at the interface of zircon with silicon, we analyze the
foreseeable effect of the defects on the electronic properties 
of zircon used as gate in metal-oxide-semiconductor devices.

\end{abstract}


\maketitle

\section{Introduction}

Its ability to accommodate actinides, its high resistance to 
corrosion and low thermal conductivity makes zircon a candidate 
ceramic for immobilization of nuclear waste\cite{Weber98,Ewing95}.  
The effect of radiation damage in these ceramics has to be
studied in order to control possible leaching, creep and fatigue in 
the material.  In radiation damaged samples, an anisotropic volume 
swelling of up to 5\% was observed in the crystalline phase.  
When the damage is not too large, there is a coexistence between 
amorphous domains and the crystalline phase.   The former is due to
collision cascades caused by the recoil particles, while the 
$\alpha$-particles produce point defects in the remaining crystalline phase.  
X-ray diffraction experiments show that this crystalline phase 
expands by up to $\sim$1.5\% along the ab-axis, and $\sim$2\% 
along the c-axis\cite{Holland55}.  
It is known that high concentrations of point defects strongly affect 
the structural properties of crystalline zircon\cite{Rios00}.  
The accumulation of charged defects produce electric fields 
that would also change the kinetics of accumulation and diffusivity.

In a different context, the high dielectric permittivity of zircon 
has recently generated considerable interest in the electronic industry.
The reduction of the size of metal-oxide-semiconductor transistors requires
the use of a gate dielectric with high electric permittivity (high-$\kappa$) 
in order to avoid the leakage, coming from direct tunnelling of carriers 
through the potential barrier of the insulator.  Alternatives to SiO$_2$ 
are being sought.  Oxides and silicates of transition metals such as Zr 
and Hf have been stabilised in contact with Si up to high temperatures.  
The silicates are specially promising because they form an interface 
with the silicon that is chemically similar to the SiO$_2$-Si 
interface\cite{Wilk00}.  The permittivity of amorphous ZrSi$_x$O$_y$ 
silicates increases as the concentration of Zr increases, but it also 
reduces the band gap of the material.  This reduction of the insulator
band gap degrades the potential barrier for tunnelling, therefore 
increasing the leakage.
Low concentrations of point defects can affect the electrical 
properties of the material, and hence have a significant role 
in the performance of zircon as a gate dielectric.
Electron and hole trapping by defects may change
the electronic structure in the band gap, and affect the leakage 
current.  

The importance of zircon for nuclear waste immobilization, where
defect accumulation in radiation-damaged samples affects the 
crystalline structure, and as an alternative 
gate dielectric, where low concentrations of defects can affect 
the device performance, motivates the study of both the high 
concentration and the low concentration limits.
Previous theoretical studies of point defects in zircon were 
focused on the structural properties, and on energetics of the 
neutral defects, showing that oxygen interstitial is the most stable 
one, with a non negligible concentration expected at thermal 
equilibrium\cite{Crocombette99}.  
In the present work, we use first principles electronic structure 
calculations to study the influence of a variety neutral and
charged point defects on the electronic structure of ZrSiO$_4$.  
The formation energies will depend on the chemical environment, 
and can be obtained as a function of the chemical potentials of 
the electrons and of the atomic species present in the system.  
The electron and hole affinities, and the ionization energies are 
important parameters that determine the charge state of the defect, 
and the capacity of trapping electrons (or holes) in the interface 
with silicon in electronic devices.

\section{Method}

\subsection{Details of the calculations}
Zircon (ZrSiO$_4$) crystallises in the tetragonal $I4_1/amd$ space group. 
The structure consists of alternating SiO$_4$ tetrahedra 
and ZrO$_8$ triangular dodecadeltahedra edge-sharing and 
forming chains parallel to the crystallographic $c$ axis.  
A body-centred unit cell can be chosen, 
containing four formula units.  The structure is fully described with four 
parameters (see Table \ref{lattice}): the lattice constants $a$ 
and $c$, and the internal parameters $y$ and $z$ that define the 
positions of oxygen atoms in the $16h$ sites. 
Calculations of the electronic properties of ZrSiO$_4$ are performed with
the self-consistent {\it ab initio} {\sc siesta} method,\cite{siesta1,siesta2}
 using Density Functional Theory (DFT)\cite{HK,KS} within the Local Density
Approximation (LDA)\cite{ca} and separable norm-conserving 
pseudopotentials.\cite{tm2,kb}  The valence wave functions are expanded
in linear combinations of strictly localized pseudoatomic numerical 
orbitals\cite{basis}.  Technical details of the pseudopotentials and 
the basis set used are described in reference \onlinecite{swelling}.
The relaxed parameters calculated for bulk zircon are in good agreement 
with experimental values (Table \ref{lattice}).  

\begin{table}[b]
\caption{Calculated structural parameters for crystalline ZrSiO$_4$
compared to experimental values. }
\begin{ruledtabular}
\begin{tabular*}{8cm}{l@{\extracolsep{\fill}}rrr}
       &  This work    &   Ref. [\onlinecite{Rignanese01}]
                                  & Expt. [\onlinecite{Robinson71}] \\
\hline
Volume \\
 $V$ (\AA$^3$)  &  129   &   127    &  131       \\
Lattice parameters \\
 $a$ (\AA)      &  6.59  &   6.54   &  6.61      \\
 $c$ (\AA)      &  5.96  &   5.92   &  5.98      \\
 $y$            & 0.068  &   0.064  &  0.066    \\
 $z$            & 0.184  &   0.194  &  0.195    \\
Interatomic distances \\
Si-O (\AA)      & 1.625  &   1.61   &  1.622    \\
Zr-O (\AA)      & 2.125  &  2.10    &  2.131    \\
                & 2.247  &  2.24    &  2.268    \\
Zr-Si(\AA)      & 2.979  &          &  2.991    \\
                & 3.617  &          &  3.626    \\
Bond angles \\
O-Si-O          & 96.16$^o$  &  97$^o$ & 97.0$^o$   \\
{O-Si-O}        & 116.50$^o$ & 116$^o$ & 116.06$^o$ \\
\end{tabular*}
\end{ruledtabular}
\label{lattice}
\end{table}

To simulate the defect structures, the host crystal is represented 
by a supercell generated by repetition of the conventional 
unit cell (4 formula units).  The point defects are then introduced 
inside this supercell (adding atoms for interstitials, removing 
atoms for vacancies, etc), that have to be large enough to describe 
the sought defect concentrations.  
We use host supercells with 24, 48, 96, and 192 atoms.  The supercell 
with 48 atoms is a 1$\times$1$\times$2 (repetition of the tetragonal 
cell along the $c$ axis), the one with 96 is a 2$\times$2$\times$1 
(repetition along the $a$ and $b$ axis), and the one
with 192 is a cell 2$\times$2$\times$2.
In this way, different concentrations of defects can be simulated by 
changing the number of repetitions in the cell.  
The unphysical divergence in the energy coming from the long range
Coulomb interactions of a periodicly repeated charged defect,
is compensated by a uniform electron-charge neutralising
background.~\cite{Payne}
In ref.[\onlinecite{swelling}] we studied the effect of high 
concentrations of intrinsic point defects in the structure of zircon and
we allowed for lattice relaxation of the supercells.  The resulting 
lattice parameters are used to simulate the limit of high concentration 
of defects.  
On the other side, the 192 supercell with the lattice
parameters of the perfect crystal structure is used for 
studying the electronic properties at low concentrations 
(isolated defects).

\subsection{Defect formation energies}
The formation energy of a defect $\alpha$ in charge state $q$ is a function 
of both the electron chemical potential $\mu_e$, and the chemical 
potentials of the species involved in the defect:
\begin{equation}
\Omega_f(\alpha,q)=E(\alpha,q)-\sum_i n_i\mu_i + q(\mu_e+E_v)
\end{equation}
\noindent 
In this expression, $E(\alpha,q)$ represents the energy of the 
defective supercell containing $n_i$ atoms of species $i$ with 
chemical potential $\mu_i$, at T=0 (we neglect entropic contributions).  
The Fermi level $\mu_e$ is measured 
relative to the top of the valence band, $E_v$.  This valence-band 
top may be shifted from the bulk value when a charged defect is 
introduced, because the potential in periodic boundary conditions 
is determined only up to a constant.  To compare energies the band 
structures of the perfect and defective supercells have to be lined 
up. We take this shift to be given by the difference in the potential 
far away from the defect in the defect cell and in the bulk 
system.~\cite{Laks92}

When the formation energies of different defects are compared, it 
should be done as a function of the chemical potentials and Fermi 
level.  The chemical potential for each element is specified by a
reference state, and it is then assumed that there is a thermodynamic 
equilibrium between the reference system (that acts as the reservoir 
of atoms where the interstitials come from and where atoms go to) 
and the zircon crystal.  
There are some thermodynamic limits to the chemical potentials: if
we take the values of bulk silicon and zirconium, and the 
molecular form of oxygen as origin for our chemical potentials, then  
\{$\mu_{\text Zr},\mu_{\text Si},\mu_{\text O}$\} are bound by: \\
(i) the values that make ZrSiO$_4$ stable, 
\begin{eqnarray}
&\mu_{\text Zr}+\mu_{\text Si}+4\mu_{\text O} = \Delta E_f({\text ZrSiO}_4) 
\label{EZrSiO4}
\end{eqnarray}
(ii) the values that cause precipitation of its constituents,
\begin{eqnarray}
&\mu_{\text Zr} \le 0, \mu_{\text Si} \le 0, \mu_{\text O} \le 0  
\end{eqnarray}
(iii) and the values that cause formation of the oxides:
\begin{eqnarray}
&\mu_{\text Zr}+2\mu_{\text O} \le \Delta E_f({\text ZrO}_2) \\
\label{EZrO2}
&\mu_{\text Si}+2\mu_{\text O} \le \Delta E_f({\text SiO}_2) 
\label{ESiO2}
\end{eqnarray}
where $\Delta E_f$ is the generalised formation free energy of the 
corresponding solid compound relative to bulk silicon, 
bulk zirconium and molecular oxygen (see table \ref{enthalpies}).  

\begin{table}[t]
\caption{Calculated formation free energies (in eV) 
for ZrSiO$_4$, ZrO$_2$, and SiO$_2$. }
\begin{tabular*}{8cm}{c@{\extracolsep{\fill}}cc}
\hline
\hline
Constituents & This work & Expt.\\
\hline
Zr + O$_2$ $\longrightarrow$ ZrO$_2$ & -12.1 & -11.5\cite{Fiorentini02} \\
Si + O$_2$ $\longrightarrow$ SiO$_2$ & -9.6 & -9.8\cite{Chase98} \\
Zr + Si + 2O$_2$ $\longrightarrow$ ZrSiO$_4$ & -22.3 & -20.9\cite{Chase98} \\
\hline
\hline
\end{tabular*}
\label{enthalpies}
\end{table}

\begin{figure}[b]
\includegraphics[scale=0.60]{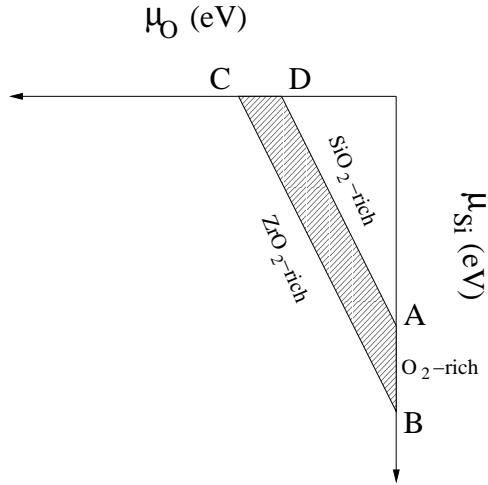}
\caption{Calculated stability quadrangle for the Zr-Si-O system in the 
\{$\mu_{\text O},\mu_{\text Si}$\} plane.  The dashed zone represents
the region of stability of zircon.  The lines DA and CB 
are determined by the equalities in equations (4) and (\ref{ESiO2}) 
respectively.  The lines AB and CD correspond to O$_2$-rich and
Si-rich environments.}
\label{triangle}
\end{figure}

Relation (\ref{EZrSiO4}) can be used to reduce the 
dependence of the defect formation energy on just a pair of atomic 
chemical potentials.  Choosing $\{\mu_{\text O},\mu_{\text Si}\}$ 
we would have the ``stability quadrangle'' shown in figure \ref{triangle}, 
where relations (4) and (5) determine the ZrO$_2$-rich and
SiO$_2$-rich environments, and the shaded region corresponds to 
values of $\{\mu_{\text O},\mu_{\text Si}\}$ that satisfy all the 
conditions (\ref{EZrSiO4}-\ref{ESiO2}).
Notice that for the study of ZrSiO$_4$ thin films grown over silicon, 
this particular choice is more convenient than 
$\{\mu_{\text Zr},\mu_{\text Si}\}$ or $\{\mu_{\text O},\mu_{\text Zr}\}$.

The Fermi level, $\mu_e$ is bound between the valence-band top, 
and the conduction-band bottom.  
\begin{eqnarray}
&0\le \mu_e \le E_g
\end{eqnarray}
The band gap, $E_g$, can be calculated as the difference of the 
energies of the system with N$-$1, N, and N$+$1 electrons\cite{Pantelides74}:
\begin{eqnarray}
E_g &=& [E(zircon,-1)-E(zircon,0)] - \\
    && [E(zircon,0)-E(zircon,+1)]
\end{eqnarray}
In this way, we obtain that the band gap is 5.12 eV (it can be compared 
with the value of 4.91 eV obtained from the calculated electronic 
band structure).  The experimental value is about 6.0 eV\cite{Robertson00}, 
but is well known that DFT (GGA and also LDA) 
underestimates $E_g$.  The difference ($\kappa$) between theoretical 
and experimental gap is the main source of inaccuracy for the defect 
levels.

\subsection{Energy Levels}
The positions of the defect levels with respect
to the bottom of the conduction band are useful for the study of 
photo-stimulated and thermo-stimulated processes.  The electron (hole)
affinity, $\chi_e(\alpha,q)$ [$\chi_h(\alpha,q)$], is defined as the 
energy gained when a free electron (hole) 
from the bottom of the conduction band (top of the valence band) is 
trapped at the defect.  It is computed by comparing energies of 
systems with the same number of electrons.  
The difference $\kappa$ between the experimental and theoretical 
energy gap can be used to correct the electron affinity~\cite{Foster02},

\begin{eqnarray}
\nonumber
\chi_e(\alpha,q)=&[E(\alpha,q)+E({\text{ZrSiO}}_4,-)]\\
-&[E(\alpha,q-1)+E({\text{ZrSiO}}_4,0)] +\kappa
\label{eq:chie}
\end{eqnarray}
\begin{eqnarray}
\nonumber
\chi_h(\alpha,q)&=&[E(\alpha,q)+E({\text{ZrSiO}}_4,+)]\\
&-&[E(\alpha,q+1)+E({\text{ZrSiO}}_4,0)]
\label{eq:chih}
\end{eqnarray}
\noindent 

With these definitions, we have that 
$\chi_e^(\alpha,q+1)+\chi_h(\alpha,q)=E_g^{exp.}$
We will consider the relaxed electron and hole affinities, that 
include the lattice relaxation after the electron/hole trapping.  
This method is approximate and the main source of errors come 
from the underestimated band gap in DFT calculations.

The defect transition energy $E_\alpha(q/q')$ defined as\cite{Zunger01}
\begin{equation}
E_\alpha(q/q') = [\Omega_f(\alpha,q) - \Omega_f(\alpha,q')]/(q'-q)
\end{equation}
determines the energy required to change the charge state of a defect 
from $q$ to $q'$.  The charge state of the defect will be $q$ if the 
Fermi level $\mu_e$ is below $E_\alpha(q/q')$, and will be $q'$ if $\mu_e$ 
is above $E_\alpha(q/q')$.

\section{Results}

We have considered a variety of neutral and charged intrinsic 
defects, including interstitials (X$_i$) and vacancies (V$_X$) of the three 
elements ($X=$Si,O,Zr) present in zircon, Zr and Si anti-site 
defects (Zr$_{\text Si}$, and Si$_{\text Zr}$), Frenkel pairs 
(X$_{\text FP}$), and other combinations of interstitials and 
vacancies.  The description of the structures of these
defects is given in Ref.~[\onlinecite{swelling}] for the high 
concentration limit, and are essentially the same for the diluted 
case.

\begin{figure*}[t]
\includegraphics[scale=0.72]{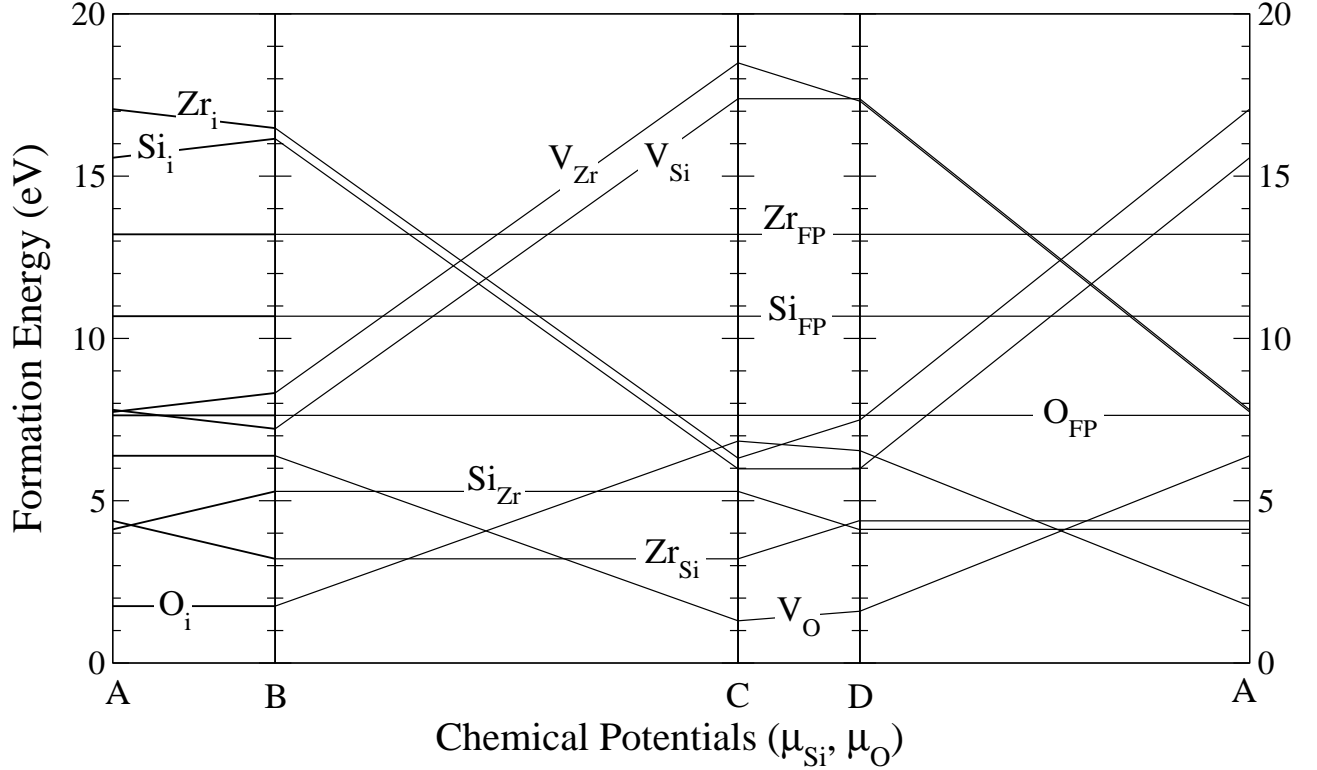}
\caption{Formation energies for point defects in ZrSiO$_4$, as a function
of the atomic chemical potentials. The labels A, B, C, and D 
correspond to the points defined in Fig.\ref{triangle}. 
Results for the high concentration limit are shown, 
though the appearance is essentially unchanged in the 
low concentration limit.}
\label{formation-at}
\end{figure*}

The presence of high concentration of point defects in zircon produces a 
strong modification of the lattice parameters.  The most important 
distortions are produced by interstitials of Si and Zr, and for the 
antisite Zr$_{\text Si}$.  For these defects, we reported\cite{swelling} 
variations in the lattice constants of the order of 1.5\% for the limit of 
high concentration studied ($2\times10^{21}$defects/cm$^{3}$).
Figure \ref{formation-at} shows the energies as a function of $\mu_{\text O}$ 
and $\mu_{\text Si}$, along the directions 
$A\rightarrow B \rightarrow C \rightarrow D$ defined in Fig.\ref{triangle}.
According to this figure, interstitials and vacancies of oxygen, 
and the antisite Zr$_{\text Si}$ are stable defects over a broad 
range of chemical potentials, followed by the Si$_{\text Zr}$ antisite.  
The energies for incorporation of native neutral point defects in ZrSiO$_4$ 
are presented in table \ref{energy}, with the reference for the
chemical potentials given in the A point.  We consider that this 
particular position, corresponding to an O$_2$-rich and SiO$_2$-rich 
environment is appropriate to describe formation energies for both 
zircon thin-films grown on silicon, and for the crystalline phase of
radiation-damaged samples.  In the latter, molecular dynamic 
simulations~\cite{Trachenko02} 
show that the content of oxygen in the cascade regions is reduced, 
becoming a source of oxygen for the crystalline regions.  Traces of
polymerisation observed in nuclear magnetic resonance 
experiments~\cite{Farnan01}, tend to indicate that the formation 
of SiO$_2$-rich domains is favorable in damaged samples.

\begin{figure}[t]
\includegraphics[scale=0.62]{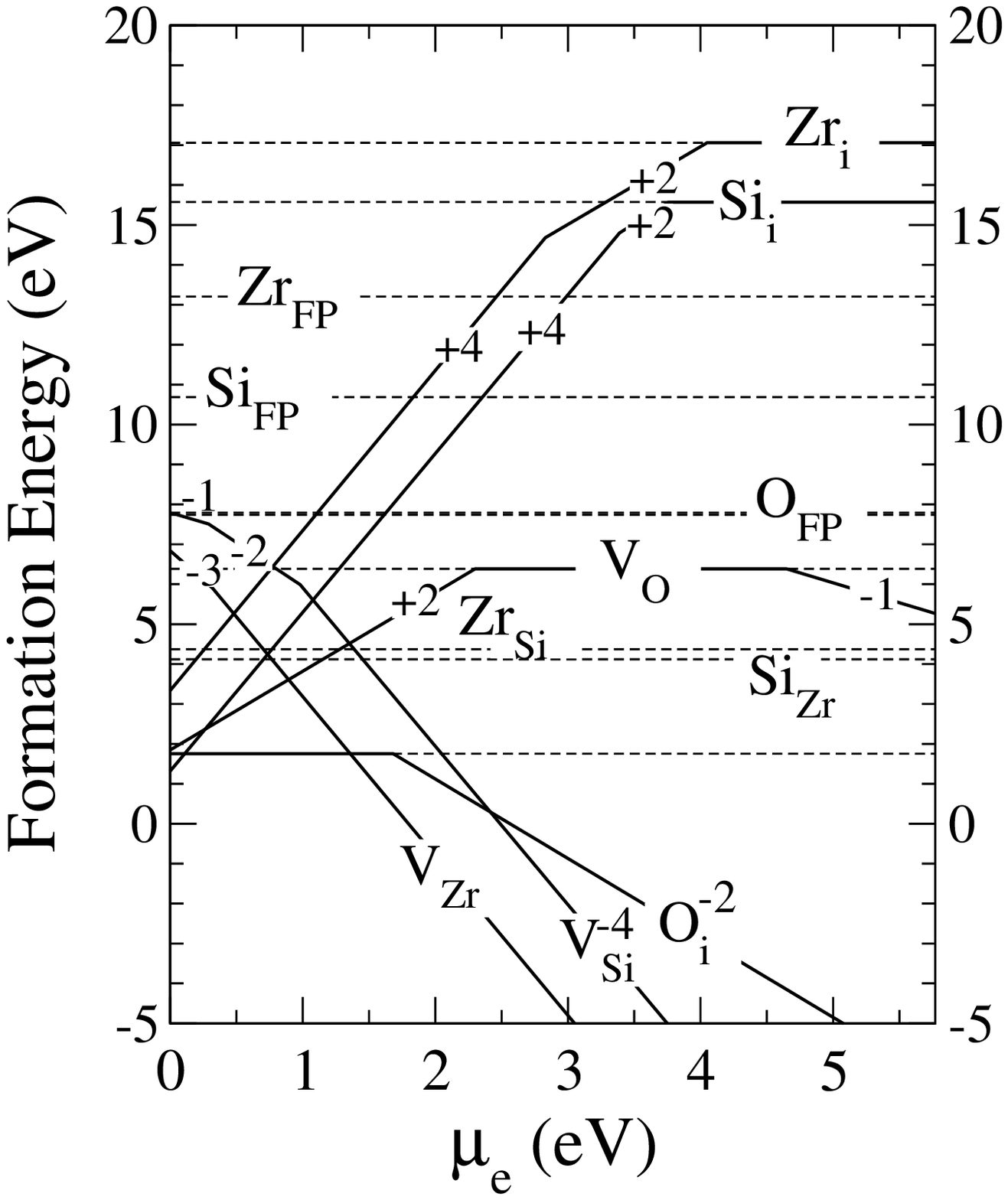}
\caption{Formation energies for point defects in ZrSiO$_4$, as a function
of the electronic chemical potential, $\mu_e$. The atomic chemical potentials
correspond to the values of the A point defined in Fig.\ref{triangle}.  
The lower energy charge states are shown for each defect, 
and the correction $\kappa$ was not considered.  
The dashed line represents the energy of the neutral configuration.}
\label{formation-e}
\end{figure}

The lattice distortion induced by the high concentration of defects 
discussed previously does not substantially affect the formation energies, 
as can be observed by the fact that the values obtained for the undistorted 
2x2x2 supercell (192 atoms) are very similar to the formation energies of 
the same defects in the distorted 1x1x2 supercell (48 atoms) shown in 
table \ref{energy}.  The small interaction between defect images is 
also reflected in the similarity of these energies.

Figure \ref{formation-e} shows the formation energies as a function of 
the electronic chemical potential $\mu_e$ for the atomic chemical 
potentials given by the point A.  For values of $\mu_e$ not too close to the
bottom of the conduction band, cation interstitials are more stable in 
positively-charged configurations.  Cation vacancies prefer 
negatively-charged states, and the same is true for the oxygen 
interstitial when $\mu_e$ is $\sim$2 eV above the top of the valence band.
In the following, we will discuss the energetic properties of each family 
of defects.

\begin{table}[t]
\caption{Defect formation energies (eV) for the neutral interstitials, 
vacancies, Frenkel pairs and antisites defects.  Results for 
different cell sizes (24, 48 or 96 atoms) 
are shown.  The chemical potentials were fixed to the values given by the 
A point (defined in Fig.\ref{triangle}), and are compared with the 
results of reference~[\onlinecite{Crocombette99}], which were obtained
for a slightly different choice of the values for the chemical potentials,
which were chemically motivated but not entirely compatible with equations (\ref{EZrSiO4}-\ref{ESiO2}).}
\begin{tabular*}{8cm}{cc@{\extracolsep{\fill}}ccccc}
\hline
 Defect $\alpha$ & 24 & 48 & 96 & 192 & Ref.~[\onlinecite{Crocombette99}] \\
\hline
\hline
O$_i$	       & 1.7  &  1.7 &  2.0 & 1.1  & 1.7       \\
Si$_i$ 	       & 14.5 & 15.6 & 15.5 & 16.4 & 17.0 \\
Zr$_i$ 	       &      & 17.1 & 17.6 & 15.9 & 18.0 \\
V$_{\text O}$  & 6.3  &  6.4 &  6.5 & 6.9  &  5.6 \\
V$_{\text Si}$ &  3.6 &  7.8 &  8.5 &      &  5.8 \\
V$_{\text Zr}$ & 7.4  &  7.7 &  7.9 &      &  5.9 \\
O$_{\text FP}$ &      &  7.6 &      &      &  7.3  \\
Si$_{\text FP}$&      & 10.7 &      &      & 22.9 \\
Zr$_{\text FP}$&      & 13.2 &      &      & 24.0 \\
Si$_{\text Zr}$&      &  4.1 &      &      &      \\
Zr$_{\text Si}$&  4.1 &  4.4 &      & 3.0  &      \\
\hline
\label{energy}
\end{tabular*}
\end{table}

\subsubsection{Oxygen related defects}
The neutral interstitial oxygen forms a ``dumbbell'' structure with 
another oxygen atom in the lattice\cite{Crocombette99,swelling}, an 
arrangement also observed in ZrO$_2$.
The structure of charged O$_i^{-}$ does not differ substantially 
from the neutral defect, but in the doubly charged state the oxygen 
displaces from the dumbbell structure, and forms a bridge between 
neighbour silicon atoms, both becoming fivefold coordinated.~\cite{swelling}  
The relaxation energy from the initial dumbbell configuration is 
of about 1.6 eV.  The energy of the neutral defect in the atomic 
structure of O$_i^{-2}$ is 2.9 eV higher in energy.  
This lowering of the energy that results from the capture of a 
second electron is known as Anderson's ``negative-U'' behaviour.~\cite{Anderson75}
It is interesting to note that the decay of two isolated O$_i^-$
centre into a O$_i^{-2}$ and a neutral O$_i$ is energetically 
favorable, with a gain of $\sim$0.4 eV (see table \ref{reactions}).

In the V$_{\text O}$, the silicon atom moves slightly towards 
the missing oxygen in its tetrahedron, and the remaining Si-O bond lengths are 
increased by 4\%.  In the charged defect V$_{\text O}^+$, the Si 
atom moves back towards its initial position. The
Si-O bonds have deviations smaller than 1\% with respect to the 
original Si-O tetrahedra.
Again, there is a negative-U behaviour, and the decay reaction 
2V$_O^+\Rightarrow$V$_O^{+2}+$V$_O$ releases an energy of 1.7 eV.  

In the case of O$_{\text FP}$, the interstitial and the vacancy 
do not interact strongly, and the final configuration is similar 
to a pair of isolated vacancy and dumbbell interstitial 
(E(O$_i$)+E(V$_{\text O}$)=8.2 eV for the supercell with 48 atoms), 
even when both are in the same original Si-O tetrahedra.  
The energy of the charged pairs (computed from the energies 
of isolated charged interstitial and vacancy) is slightly different, 
with V$_O^{+}+$O$_i^{-}$ 0.5 eV higher in energy, 
and V$_O^{+2}+$O$_i^{-2}$ 1.2 eV lower than the neutral pair, 
showing a tendency towards charge transfer between oxygen 
vacancies and interstitials.

\begin{table}[b!]
\caption{Energies for defect reactions, obtained from the formation 
energies of isolated defects.}
\begin{tabular*}{8cm}{rcl@{\extracolsep{\fill}}dd}
\hline
\multicolumn{3}{c}{ } & \multicolumn{2}{c}{Energy (eV)} \\
\multicolumn{3}{c}{ Reaction} & \multicolumn{1}{c}{high-C} & \multicolumn{1}{c}{low-C} \\
\hline
\hline
O$_i^0$ + O$_i^{2-}$& $\longrightarrow$& $2$O$_i^{-}$ & -0.4 & -1.1 \\
V$_O^0$ + V$_O^{2+}$& $\longrightarrow$& $2$V$_O^{+}$ & -1.7 & -1.0 \\
O$_i^0$ + V$_O^{0}$ & $\longrightarrow$& O$_i^{-}$ +V$_O^{+}$ & -0.5 & -0.7 \\
O$_i^-$ + V$_O^{+}$ & $\longrightarrow$& O$_i^{2-}$ +V$_O^{2+}$ & 1.7 &  1.4 \\
O$_i^0$ + V$_O^{0}$ & $\longrightarrow$& O$_i^{2-}$ +V$_O^{2+}$ & 1.2 &  0.7 \\
Si$_i^0$ + Si$_i^{+2}$ & $\longrightarrow$&  $2$Si$_i^{+}$ & -1.5 & -1.0 \\
Si$_i^0$ + Si$_i^{+4}$ & $\longrightarrow$&  $2$Si$_i^{2+}$ & 0.6 & 0.8 \\
Si$_i^{+2}$ + Si$_i^{+4}$ & $\longrightarrow$&  $2$Si$_i^{3+}$ & -2.8 & -2.6 \\
V$_{Si}^0$ + V$_{Si}^{2-}$& $\longrightarrow$& $2$V$_{Si}^{-}$ &  0.3 &     \\
V$_{Si}^0$ + V$_{Si}^{4-}$& $\longrightarrow$& $2$V$_{Si}^{2-}$ & 1.7 &     \\
Si$_i^0$ + V$_{Si}^{0}$& $\longrightarrow$& Si$_i^{+}$ +V$_{Si}^{-}$ & 3.0 & \\
Si$_i^0$ + V$_{Si}^{0}$& $\longrightarrow$& Si$_i^{2+}$ +V$_{Si}^{2-}$ & 7.2 & \\
Si$_i^0$ + V$_{Si}^{0}$& $\longrightarrow$& Si$_i^{3+}$ +V$_{Si}^{3-}$ & 8.4 & \\
Si$_i^0$ + V$_{Si}^{0}$& $\longrightarrow$& Si$_i^{4+}$ +V$_{Si}^{4-}$ &12.2 & \\
Zr$_i^0$ + Zr$_i^{+2}$ & $\longrightarrow$&  $2$Zr$_i^{+}$ & -0.6 & 0.02 \\
Zr$_i^0$ + Zr$_i^{+4}$ & $\longrightarrow$&  $2$Zr$_i^{2+}$ & 2.3 & 2.2 \\
Zr$_i^{+2}$ + Zr$_i^{+4}$ & $\longrightarrow$&  $2$Zr$_i^{3+}$ & -0.2 &  0.1 \\
V$_{Zr}^0$ + V$_{Zr}^{2-}$& $\longrightarrow$& $2$V$_{Zr}^{-}$ &  0.2 &     \\
V$_{Zr}^0$ + V$_{Zr}^{4-}$& $\longrightarrow$& $2$V$_{Zr}^{2-}$ & 1.3 &     \\
Zr$_i^0$ + V$_{Zr}^{0}$& $\longrightarrow$& Zr$_i^{+}$ +V$_{Zr}^{-}$ & 4.3 & \\
Zr$_i^0$ + V$_{Zr}^{0}$& $\longrightarrow$& Zr$_i^{2+}$ +V$_{Zr}^{2-}$ & 9.0 & \\
Zr$_i^0$ + V$_{Zr}^{0}$& $\longrightarrow$& Zr$_i^{3+}$ +V$_{Zr}^{3-}$ & 11.8 & \\
Zr$_i^0$ + V$_{Zr}^{0}$& $\longrightarrow$& Zr$_i^{4+}$ +V$_{Zr}^{4-}$ & 14.4 & \\
\hline
\end{tabular*}
\label{reactions}
\end{table}

\subsubsection{Cation related defects}
Neutral interstitials of Zr (Si) are relatively stable in a Zr-rich (Si-rich) 
environment, but in general have higher formation energies.  These 
defects are more likely to exist in their charged states X$_i^{+2}$ or 
X$_i^{+4}$ (with X=Zr or Si) than in the neutral state, depending in the
position of the Fermi level (figure \ref{formation-e} and table 
\ref{reactions}).  
Upon removal of electrons from Si$_i$, new atomic structures are obtained,
with the interstitial making oxygen atoms to approach, and neutralize the
charge around it.  The geometries for positively charged Zr-interstitials 
are basically unchanged with respect to the neutral structure.  
The combination of neutral isolated vacancies with interstitials 
would favour a charge transfer between them.
In that case, there is a strong ionic interaction between interstitial and
vacancy and a clear trend towards defect annihilation.  This 
results in relatively high formation energies for Frenkel pairs of Zr and Si.
Our values for the formation energies in these defects are a factor of 
two smaller than the values computed from the formation energies of 
the neutral interstitial and the neutral vacancy (23.4 and 25 eV for 
Si$_{FP}$ and Zr$_{FP}$ respectively).  Doing the same for the 
charged pairs (V$_{\text Zr}^{-n}+$Zr$_i^{+n}$ and 
V$_{\text Si}^{-n}+$Si$_i^{+n}$, for $n=1,2,3,4$) we obtain that the energy
decreases from $\sim$20 eV to $\sim$11 eV for the 
V$_{\text Zr}^{-4}+$Zr$_i^{+4}$ pair, and the same for
V$_{\text Si}^{-4}+$Si$_i^{+4}$.  This suggests that the Frenkel pairs,
Si$_{\text FP}$ and Zr$_{\text FP}$, that we actually simulated are 
more close to the charged configuration than to the neutral one, showing 
the strong ionic character of these defects.
Si$_{\text Zr}$ and Zr$_{\text Si}$ antisites have relatively low 
formation energies, and their neutral state is the most stable.

\subsubsection{Defect activity}
The calculated relaxed electron and hole affinities (with 
the $\kappa$ correction for the electron affinities)
are summarised in Table \ref{affinities}.  The values of the $\chi_e$ 
for oxygen interstitials are similar to the ones obtained for O in 
ZrO$_2$\cite{Foster01} and HfO$_2$\cite{Foster02} and
indicate that these defects may serve as traps for electrons from the 
conduction band.  The same is true for the Si and Zr vacancies.  
On the other hand, the high value of the hole affinities for Zr and Si
interstitials show that these defects tend to act as traps for holes 
from the valence band.

\begin{figure}[b]
\includegraphics[scale=0.35]{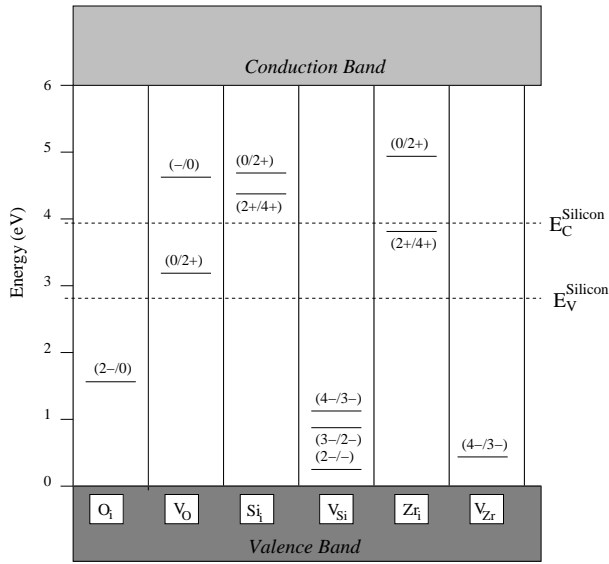}
\caption{Energy level diagram showing the defect levels in ZrSiO$_4$, and the reference of conduction and valence band edges of silicon.}
\label{levels}
\end{figure}

The presence of point defects in ZrSiO$_4$ thin films will affect the
performance of this material as an alternative gate dielectric in 
microelectronics.  Recent {\it ab initio} calculations\cite{Puthenkovilakam04} 
showed that zircon
forms an excellent interface with silicon providing adequate barriers
for both electrons and holes (the band alignment is symmetric, with 
valence band offset of 2.78 eV and conduction band offset of 2.10 eV).
The alignment of the defect levels with the valence and conduction 
band edges will determine whether these defects can play a role in 
the conductivity properties of the barrier.  Figure \ref{levels} show 
the calculated defect transition energy levels $E(q/q')$ with the 
reference of the calculated silicon valence and conduction band edges 
presented in Ref.~[\onlinecite{Puthenkovilakam04}].  
The defect levels are relatively deep.
The acceptor levels for O$_i$, V$_{\text Si}$, and V$_{\text Zr}$ can 
trap electrons injected from the top of the silicon valence band.
The donor level $E(0/2+)$ for oxygen vacancy lies in the silicon energy 
gap and can act as hole-killer center for p-type doped silicon, 
or trap electrons for n-type doped silicon.  The $E(2+/4+)$ level 
for Zr interstitial practically resonates with the bottom of the silicon
conduction band, and would act as a shallow donor.  
Donor levels for interstitial Si lie just above the silicon 
conduction band minimum and would readily produce electrons into the device.

\begin{table}[t]
\caption{Electron and hole affinities ($\chi_e$ and $\chi_h$), as
defined in equations (\ref{eq:chie}) and (\ref{eq:chih}), for the
interstitials and vacancies in different charge states. The energies
are given in eV.}
\begin{tabular*}{5.5cm}{l@{\extracolsep{\fill}}dddd}
\hline
\hline
 Defect & \multicolumn{2}{c}{$\chi_e$} & \multicolumn{2}{c}{$\chi_h$} \\
              & \multicolumn{1}{c}{high-C} & \multicolumn{1}{c}{low-C} & \multicolumn{1}{c}{high-C} & \multicolumn{1}{c}{low-C} \\
\hline
O$_i^{2-}$    &        &       &  1.8   & 1.4   \\
O$_i^{-}$     &   4.1  &  4.6  &  2.3   & 2.5   \\
O$_i$         &   3.7  &  3.5  &  0.3   &       \\
O$_i^{+}$     &   5.7  &       &  0.3   &       \\
V$_O^{-}$     &        &       &  5.0   &       \\
V$_{O}$       &   1.0  &       &  1.8   & 1.8   \\
V$_O^{+}$     &   4.2  &  4.2  &  3.5   & 2.8   \\
V$_O^{2+}$    &   2.5  &  3.2  &        &       \\

Si$_i$        &        &       &  3.4   & 3.7   \\
Si$_i^+$      &   2.6  &  2.3  &  4.9   & 4.7   \\
Si$_i^{2+}$   &   1.1  &  1.2  &  2.4   & 2.6   \\
Si$_i^{3+}$   &   3.6  &  3.4  &  5.2   & 5.1   \\
Si$_i^{4+}$   &   0.7  &  0.8  &        &       \\

V$_{Si}^{4-}$ &        &       &  1.5   &       \\
V$_{Si}^{3-}$ &   4.5  &       &  1.3   &       \\
V$_{Si}^{2-}$ &   4.7  &       &  0.6   &       \\
V$_{Si}^-$    &   5.4  &       &  0.4   &       \\
V$_{Si}$      &   5.6  &       &  0.1   &       \\

Zr$_i$        &        &       &  4.1   & 4.1   \\
Zr$_i^+$      &   1.9  & 1.9   &  4.7   & 4.1   \\
Zr$_i^{2+}$   &   1.3  & 1.9   &  3.2   & 3.0   \\
Zr$_i^{3+}$   &   2.8  & 2.9   &  3.4   & 2.9   \\
Zr$_i^{4+}$   &   2.6  & 3.1   &        &       \\

V$_{Zr}^{4-}$ &        &       &  0.7   &       \\
V$_{Zr}^{3-}$ &   5.3  &       &  0.4   &       \\
V$_{Zr}^{2-}$ &   5.6  &       &  0.0   &       \\
V$_{Zr}^-$    &   5.9  &       & -0.2   &       \\
V$_{Zr}$      &   6.3  &       & -0.3   &       \\

Zr$_{Si}$     &   0.6  & 0.9   &  0.5   & 0.4   \\
\hline
\hline
\label{affinities}
\end{tabular*}
\end{table}

\section{Conclusions}

We have used first principles simulations to study the energetics of 
high and low concentrations of point defects in zircon.  
The effect of volume swelling induced by
high concentration of defects in the formation energies is of the order
of tenths of eV, and is not affecting the relative stability of
the intrinsic point defects.  We have seen that interstitials and vacancies
of oxygen and the Zr$_{\text Si}$ and Si$_{\text Zr}$ anti-sites 
are the most stable defects.  
There is a strong tendency towards ionization of the defects, and a 
negative-U behaviour was observed for some defects in this material.  
Interstitials of oxygen, and the vacancies of Zr and Si may act as 
traps for electrons, while vacancies of oxygen and cation interstitials 
would serve as traps for holes. 
The strong ionic character of cation defects is also shown by the charge 
transfer between vacancies and interstitials to form neutral Frenkel pairs.
The deep defect levels induced in the gap will increase the leakage and
reduce the performance of zircon as a gate dielectric.
The presence of high concentration of charged defects in zircon for 
nuclear waste immobilization will create internal electric fields that 
can affect the kinetics of impurity mobility and their effect has to be 
included in any modelization of these materials.  

\acknowledgments
This work was supported by British Nuclear Fuels (BNFL) and NERC.
We would like to thank E. K. H. Salje, M. T. Dove, K. Trachenko,
M. Yang, S. Rios, I. Farnan for helpful discussions on zircon.

\bibliography{defects}

\end{document}